\documentclass[11pt,fleqn]{article}

\usepackage{amsmath,amssymb,graphicx}
\usepackage{verbatim}
\begin{document}
\sloppy
\newtheorem{axiom}{Axiom}[section]
\newtheorem{claim}[axiom]{Claim}
\newtheorem{conjecture}[axiom]{Conjecture}
\newtheorem{corollary}[axiom]{Corollary}
\newtheorem{definition}[axiom]{Definition}
\newtheorem{example}[axiom]{Example}
\newtheorem{fact}[axiom]{Fact}
\newtheorem{lemma}[axiom]{Lemma}
\newtheorem{observation}[axiom]{Observation}
\newtheorem{proposition}[axiom]{Proposition}
\newtheorem{theorem}[axiom]{Theorem}

\renewcommand{\topfraction}{1.0}
\renewcommand{\bottomfraction}{1.0}

\newcommand{\proof}{\emph{Proof.}\ \ }
\newcommand{\qed}{~~$\Box$}
\newcommand{\rz}{{\mathbb{R}}}
\newcommand{\nz}{{\mathbb{N}}}
\newcommand{\zz}{{\mathbb{Z}}}
\newcommand{\eps}{\varepsilon}
\newcommand{\cei}[1]{\lceil #1\rceil}
\newcommand{\flo}[1]{\left\lfloor #1\right\rfloor}
\newcommand{\seq}[1]{\langle #1\rangle}
\newcommand{\p}{\star}

\title{{\bf Travelling salesman paths on Demidenko matrices}}
\author{
Eranda \c{C}ela\thanks{{\tt cela@opt.math.tu-graz.ac.at}.
Department of Discrete Mathematics, TU Graz, Austria}
\and
Vladimir G. Deineko\thanks{{\tt Vladimir.Deineko@wbs.ac.uk}.
Warwick Business School, University of Warwick, United Kingdom}
\and
Gerhard J. Woeginger\thanks{{\tt woeginger@algo.rwth-aachen.de}.
Department of Computer Science, RWTH Aachen, Germany}
}
\date{}
\maketitle

\begin{abstract}
In the path version of the Travelling Salesman Problem (Path-TSP), a salesman is looking 
for the shortest Hamiltonian path through a set of $n$ cities. 
The salesman has to start his journey at a given city $s$, visit every city exactly once, 
and finally end his trip at another given city $t$.

In this paper we identify a new polynomially solvable case of the Path-TSP where the 
distance matrix of the cities is a so-called Demidenko matrix. 
We identify a number of crucial combinatorial properties of the optimal solution, and 
we design a dynamic program with time complexity $O(n^6)$. 

\end{abstract}

\medskip
\section{Introduction}
The travelling salesman problem (TSP) is one of the best studied problems in operational research. 
This is not only due to the numerous appearances of the TSP in various practical applications, 
but also due to its pivotal role in developing and testing new research methods.
We refer the reader to the books by
Applegate, Bixby, Chv\'{a}tal \& Cook \cite{Applegate},
Gutin \& Punnen \cite{Gutin}, and
Lawler, Lenstra, Rinnooy Kan \& Shmoys \cite{TSP-book}
for a wealth of information on these issues.
We also emphasize the role of the TSP in education and in the popularization of science. 
There is hardly any text book on operational research where the TSP would not be mentioned.
The book \cite{Cook} of Cook is an excellent example of how the TSP 
is used in the popularization of science. 

An instance of the TSP consists of $n$ cities together with an $n\times n$ symmetric distance matrix 
$C=(c_{ij})$ that specifies the distance $c_{ij}$ between any pair $i$ and $j$ of cities.
The objective in the TSP is to find a shortest closed route which visits each city exactly once; 
such a closed route is called a TSP tour. 
In the \emph{Path-TSP} the instance also specifies two cities $s$ and $t$ (with $s\ne t$), and the 
goal is to find a shortest route starting at city~$s$, ending at city $t$, and visiting all the 
other cities exactly once.
Both the TSP and the Path-TSP are NP-hard to solve exactly (see for instance Garey \& Johnson \cite{GJ}), 
and both problems are APX-hard to approximate (Papadimitriou \& Yannakakis \cite{PaYa93}; 
Zenklusen \cite{Zenklusen}). 
These intractability results hold even in the metric case, where the distances between the cities are 
non-negative and satisfy the triangle inequality. 
Given the intractability of TSP and Path-TSP, the characterization of tractable special cases is of 
obvious interest and forms a well-established and vivid branch of research.

\paragraph{Polynomially solvable cases of the TSP.}
The literature contains an impressive number of polynomially solvable cases for the classical TSP, 
as certified by the surveys of 
Burkard \& al~\cite{BSurv},
Deineko, Klinz, Tiskin \& Woeginger \cite{DeKlTiWo2014},
Gilmore, Lawler \& Shmoys \cite{GLS}, and 
Kabadi \cite{Kabadi}, 
and the references therein. 
We now briefly discuss three tractable cases of the TSP that are relevant for the current paper.

In the so-called \emph{Convex-Euclidean TSP}, the cities are points in the Euclidean plane which 
are all located on the boundary of their convex hull.
A folklore result says that the optimal tour in the Convex-Euclidean TSP is the cyclic walk 
along the convex hull, taken either in clockwise or in counter-clockwise direction.
If we number the cities as $1,2,\ldots,n$ in clockwise order along the convex hull, then
the cities satisfy the so-called \emph{quadrangle inequalities}:
\begin{align}
c_{ij}+c_{k\ell} ~\le~ c_{j\ell}+c_{ik} &\text{\qquad for all~} 1\le i<j<k<\ell\le n  \label{eq:kalm1} \\[0.5ex]
c_{i\ell}+c_{jk} ~\le~ c_{j\ell}+c_{ik} &\text{\qquad for all~} 1\le i<j<k<\ell\le n  \label{eq:kalm2}
\end{align}
These quadrangle inequalities state the fact that in a convex quadrangle, the total length 
of two opposing sides is less or equal to the total length of the two diagonals.
Kalmanson \cite{Kalmanson1975} observed that whenever a TSP instance satisfies these quadrangle 
inequalities \eqref{eq:kalm1}--\eqref{eq:kalm2}, the tour $1,2,\ldots,n$ is a shortest TSP tour.
This extends the tractability of the Convex-Euclidean TSP to the tractability of the TSP on so-called 
\emph{Kalmanson matrices}, where the distances satisfy \eqref{eq:kalm1}--\eqref{eq:kalm2}.
We stress that the class of Kalmanson distance matrices is large and goes far beyond the Convex-Euclidean case:
Consider for instance a rooted ordered tree with non-negative edge lengths, place a city in each of the
leaves, and number the cities from left to right.
Then the shortest path distances $c_{ij}$ between cities $i$ and $j$ determine a Kalmanson matrix, as 
the inequalities \eqref{eq:kalm1} and \eqref{eq:kalm2} can easily be verified for any quadruple of leaves.
Finally, we mention the TSP on \emph{Demidenko matrices}, where the distances satisfy
\begin{align}
c_{ij}+c_{k\ell} ~\le~ c_{j\ell}+c_{ik} &\text{\qquad for all~} 1\le i<j<k<\ell\le n.  \label{eq:demi} 
\end{align}
Note that condition \eqref{eq:demi} coincides with condition \eqref{eq:kalm1}. 
Hence Demidenko matrices form an obvious generalization of Kalmanson matrices, and it is known that 
this generalization is proper.
A famous result of Demidenko~\cite{Demidenko1979} shows that the TSP on Demidenko matrices is 
solvable in polynomial time.
The gist of this paragraph is that the Convex-Euclidean TSP, the TSP on Kalmanson matrices, and 
the TSP on Demidenko matrices form three tractable TSP cases of strictly increasing generality.

\begin{figure}[bht]
\unitlength=1cm
\centering
\begin{picture}(15,7.5)
\put(-0.1,2.7){\framebox[0.47\width]{ \centerline{\includegraphics[scale=0.9]{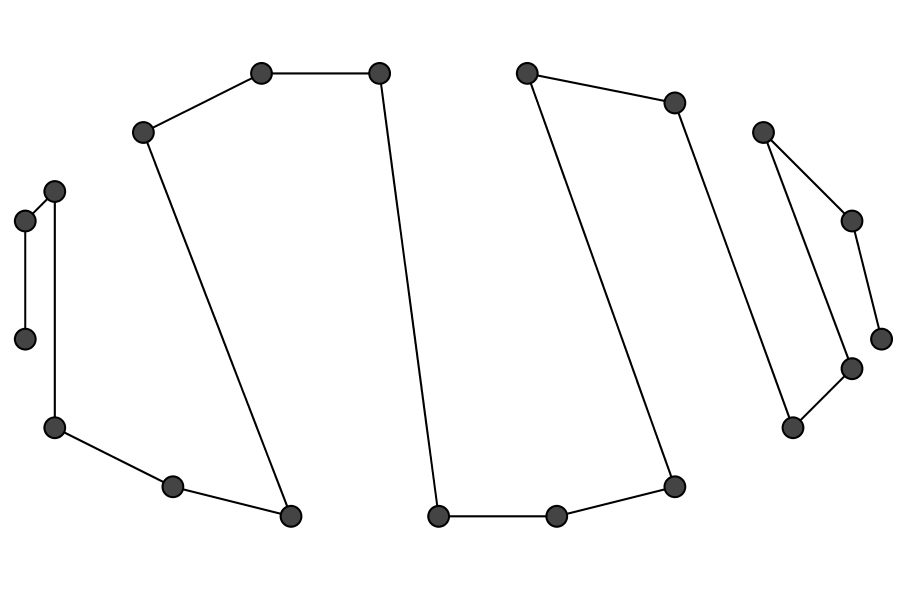}} }}
\put(3.5,2.){(a)}
\put(11,2.){(b)}
\put(7.9,2.7){\framebox[0.47\width]{
\centerline{\includegraphics[scale=0.9]{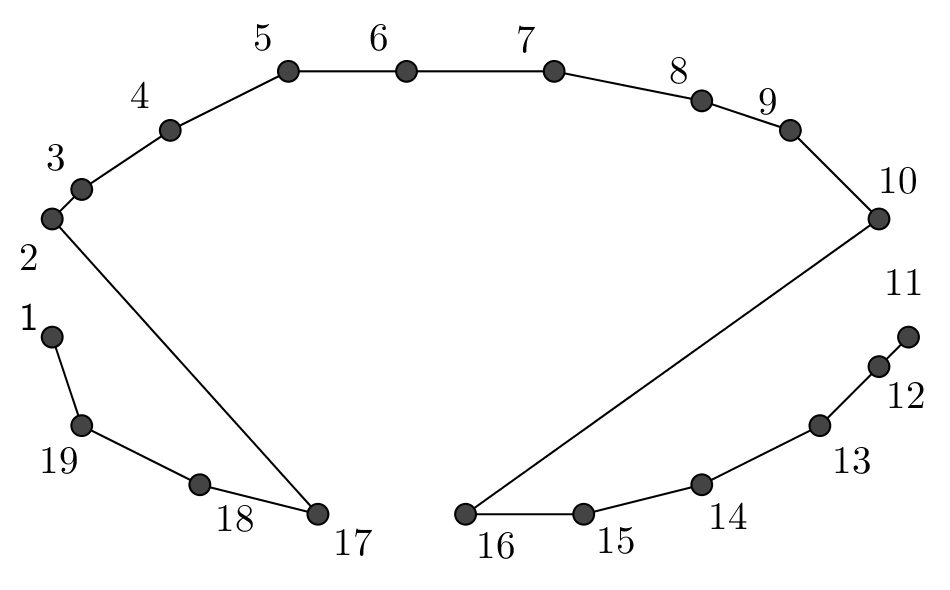}}
}}
\put(-0.1,0.5)
{\small
\begin{tabular}{|r|r|r|r|r|r|r|r|r|r|r|r|r|r|r|r|r|r|r|r|}
\hline
 &1&2&3&4&5&6&7&8&9&10&11&12&13&14&15&16&17&18&19\\
 \hline\hline
X&0&0&1&4&8&12&17&22&25&28&29&28&26&22&18&14&9&5&1\\
Y&6&10&11&13&15&15&15&14&13&10&6&5&3&1&0&0&0&1&3\\
\hline
\end{tabular}}
\end{picture}
\caption{The Path-TSP in the Convex-Euclidean case:
(a) An illustrating example from \cite{Garcia1};
(b) An optimal (1,11)-path for the set of points listed in the table.}
\label{fig:Convex1}
\end{figure}

\paragraph{Polynomially solvable cases of the Path-TSP.}
Whereas the literature on polynomially solvable cases of the classical TSP forms a rich and comprehensive 
body, we are only aware of a single result on polynomially solvable cases of the Path-TSP:
Garcia \& Tejel~\cite{Garcia1} derive an $O(n\log n)$ algorithm for the Path-TSP with $n$ cities 
on Convex-Euclidean distance matrices.
The follow-up work \cite{Garcia2} by Garcia, Jodra \& Tejel uses sophisticated search techniques 
to improve the time complexity to linear time $O(n)$. 
Figure~\ref{fig:Convex1} provides an example for the Convex-Euclidean Path-TSP, which is taken from 
Figure~3 in~\cite{Garcia1}; this example illustrates the diverse and manifold shapes of TSP-paths 
under Convex-Euclidean distances.

Now by looking deeper into the papers \cite{Garcia1,Garcia2} and by carefully analyzing the flow 
of arguments, one realizes that the approach does not exploit any geometric property of the 
Convex-Euclidean case that would go beyond the quadrangle inequalities \eqref{eq:kalm1}--\eqref{eq:kalm2}.
In other words, the arguments in \cite{Garcia1,Garcia2} do not only settle the Path-TSP on 
Convex-Euclidean distance matrices, but they do also yield (without additional effort, and without 
changing a single letter) a polynomial time solution for the Path-TSP on Kalmanson matrices.
Note that this step from Convex-Euclidean matrices to Kalmanson matrices for the Path-TSP runs perfectly 
in parallel with the step from Convex-Euclidean matrices to Kalmanson matrices for the classical TSP.

\paragraph{Contribution and organization of the paper.} 
In this paper we take the logically next step in this line of research and show that the Path-TSP 
is polynomially solvable on Demidenko matrices.
This substantially generalizes and extends the results in \cite{Garcia1,Garcia2} for the Path-TSP
on Convex-Euclidean matrices and Kalmanson matrices.
We first analyze the combinatorial structure of an optimal TSP-path on Demidenko matrices, and prove
that there always exists an optimal solution of a certain strongly restricted and nicely structured form.
Then we show that we can optimize in polynomial time over the TSP-paths of that nicely structured form.
By combining these results, we get our polynomial time result for Demidenko matrices.

The remainder of the paper is organized as follows. 
In Section~\ref{sec:DefNot} we summarize definitions and notations related to paths and tours as well as 
concrete matrix classes of relevance in this paper. 
In Section~\ref{sec:structure} we introduce the concept of {\sl forbidden pairs of
arcs} and show that for arbitrary cities $s$ and $t$ there always exists an
optimal $(s,t)$-TSP-path which does not contain forbidden pairs of arcs. Then
we show some implications of  this result in terms of  the solution of the Path-TSP
for $s=1$ and $t=n$,  and of structural properties of the optimal solution of the
problem in the more general case with $s=1$ and arbitrary $t$.   
Section~\ref{sec:s=1} shows how to  exploit the findings presented in
Section~\ref{sec:structure} to efficiently solve the Path-TSP for $s=1$ and 
for arbitrary $t$ by dynamic programming. 
Finally, Section~\ref{sec:s=general} shows how to efficiently solve the Path-TSP in the most 
general case, where there are no restrictions on $s$ and $t$. 
Section~\ref{sec:remarks} concludes the paper with some final remarks.
\section{Definitions, notations and preliminaries}
\label{sec:DefNot}
In this section we summarize various definitions and notations that will be used
throughout the rest of the paper.

\subsection{Paths and tours}\label{DefNot:path}
We consider a set of $n$ cities with a symmetric $n\times n$ distance matrix $C=(c_{ij})$.
We use the notation $\overline{a,b}$ for the set $\{a,a+1,\ldots,b\}$ of all integers between 
$a$ and $b$, for any two integers $a$, and $b$ with $a \le b$. 

A \emph{tour} $\tau$ visiting $k$ cities, $3\le k\le n$, is a
sequence $\tau=\seq{\tau_1,\tau_2,\ldots,\tau_{k},\tau_{k+1}}$ with $\tau_i\in
\overline{1,n}$, $\tau_{k+1}=\tau_1$ and $\tau_1<\tau_i$, for all $i\in \overline{2,k}$, 
such that every city from $\overline{1,n}$ except for $\tau_1$ is contained at most once in the sequence. 
We say that the tour $\tau$ visits the cities $\tau_1$, $\ldots$, $\tau_k$. 
A tour $\tau=\seq{\tau_1=1,\tau_2,\ldots, \tau_{n},1}$ visiting all $n$ cities is called \emph{a TSP tour}. 
A tour $\tau=\seq{\tau_1,\tau_2,\ldots, \tau_k,\tau_{k+1}}$ is called \emph{pyramidal} 
if there exists an $l\in \overline{1,k}$ such that $\tau_1<\ldots<\tau_l$, and $\tau_l>\ldots >\tau_{k}$ hold.
A TSP tour which is pyramidal is called \emph{a pyramidal TSP tour}. 

An \emph{$(s,t)$-path} visiting $k$ cities is a sequence $\tau=\seq{\tau_1=s,\tau_2,\ldots,\tau_k=t}$ 
of cities which starts at $s$, ends at $t$, and contains every city from $\{\tau_1,\ldots,
\tau_k\}\subseteq \overline{1,n}$ exactly once, where $k\in \nz$, $k\le n$. 
An \emph{$(s,t)$-TSP-path} is an \emph{$(s,t)$-path} visiting all $n$ cities. 
An $(s,t)$-path $\tau=\seq{\tau_1=s,\tau_2,\ldots,\tau_k=t}$ is called
\emph{$\lambda$-pyramidal} if there exists an index $l\in \overline{1,,k}$ such that 
$\tau_1<\tau_2<\ldots <\tau_l$ and $\tau_l>\tau_{l+1}>\ldots >\tau_k$. 
An $(s,t)$-path $\tau=\seq{\tau_1=s,\tau_2,\ldots,\tau_k=t}$ is called
\emph{$\nu$-pyramidal} if there exists an index $l\in \overline{1,,k}$ such that 
$\tau_1>\tau_2>\ldots >\tau_l$ and $\tau_l<\tau_{l+1}<\ldots <\tau_k$. 
(In the last two definitions one of the chains of inequalities would become obsolete if $l=1$ or $l=k$.)

Given a tour or an $(s,t)$-path $\tau=\seq{\tau_1,\tau_2,\ldots,\tau_{k}}$ and two indices 
$i\in \overline{1,k-1}$, $j\in \overline{2,k}$, then $\tau(\tau_i):=\tau_{i+1}$ is called 
the successor of $\tau_i$ in $\tau$ and $\tau^ {-1}(\tau_j):=\tau_{j-1}$ is called the 
predecessor of $\tau_j$ in $\tau$. 
An ordered pair $(\tau_i, \tau_{i+1}=\tau(\tau_i))$, for $i\in \overline{1,,k-1}$, is called an 
\emph{arc} in $\tau$. If $\tau_i<\tau_{i+1}$ ($\tau_i>\tau_{i+1}$), $(\tau_i, \tau_{i+1})$ is 
called an \emph{increasing arc} (\emph{decreasing arc}). 
An $(s,t)$-path such that all its arcs are increasing (decreasing)
arcs is called a \emph{monotone increasing path} (\emph{monotone decreasing path}). 
For $i \in \overline{2, k-1}$ a city $\tau_i$ in $\tau$ is called a \emph{peak} in $\tau$ if
$\tau_i>\tau_{i-1}$ and $\tau_i>\tau_{i+1}$. 
A city $\tau_i$ is called a \emph{valley} in $\tau$ if $\tau_i<\tau_{i-1}$ and $\tau_i<\tau_{i+1}$. 
Thus a peak (a valley) in an $(s,t)$-path $\tau$ is a city which is reached along an increasing 
(a decreasing) arc and is left along a decreasing (an increasing) arc. 
The first city $s$ of an $(s,t)$-path $\tau$ ($s\neq t$) is called a peak if $s<\tau(s)$ and a valley if $s>\tau(s)$. 
The last city $t$ of an $(s,t)$-path $\tau$ ($s\neq t$) is called a peak if $\tau^{-1}(t)>t$) and a valley if $\tau^{-1}(t)<t$. 
Given a tour or an $(s,t)$-path $\tau=\seq{\tau_1,\tau_2,\ldots,\tau_k}$ and two indices 
$i,j\in\overline{1,k}$, $i\neq j$, the subsequence of $\tau$ starting at city $\tau_i$
and ending at city $\tau_j$ is called the $(\tau_i,\tau_j)$-subpath of $\tau$.

Next we introduce the four quantities $E_m(i,j)$, $D_w(i,j)$, $\Lambda_m(i,p)$, and $V_w(j,q)$ for 
integers $i,j,m,w,p,q\in\overline{1, n}$, which represent the lengths of certain $\lambda$-pyramidal paths 
or $\nu$-pyra\-midal paths in the following way:
For $i,j,m \in \overline{1,n}$ with $i<j\le m$ denote by $E_m(i,j)$ the length of a shortest 
$\lambda$-pyramidal $(i,j)$-path $\tau$ visiting the cities $\{i\}\cup\overline{j,m}$. 
Notice that due to the symmetry of the distance matrix $E_m(i,j)$ is also the length of a shortest 
$\lambda$-pyra\-midal $(j,i)$-path $\tau$ visiting the above set of cities. 
Clearly in a $\lambda$-pyra\-midal $(i,j)$-path as above city $j+1$ is either the successor of $i$ or the
predecessor of $j$. 
Analogously in a pyramidal TSP-tour city $1$ is visited right before or right after city $2$. 
Thus the optimal length of a pyramidal tour equals $E(1,2)+c_{21}$. 
If $j=m$, then $E_{m}(i,m)=c_{im}$ and the corresponding path consists just of the arc $(i,m)$.

For $w,j,i \in \overline{1,n}$, $ w\le j<i$, let $D_w(i,j)$ be the length of a shortest $\nu$-pyramidal 
$(i,j)$-path which visits the cities $\overline{w,j}\cup\{i\}$ (and also the length of a shortest 
$\nu$-pyramidal $(j,i)$-path visiting the above set of cities). 
Clearly in a $\nu$-pyramidal $(i,j)$-path as above city $j-1$ is either the successor of $i$ or the
predecessor of $j$. If $j=w$, then $D_{w}(i,j)=c_{iw}$ and the corresponding path consists just of the arc $(i,w)$.

Next we introduce notations for the length of shortest $\lambda$-pyramidal paths and shortest $\nu$-pyramidal
paths which visit \emph{contiguous} sets of cities, i.e.\ sets of cities which consist of all 
cities $k$ between $i$ and $j$ for some pair of cities $i,j\in \overline{1,n}$, $i<j$.
 
For $i,p,m\in \overline{1,n}$, $i<p\le m$, let $\Lambda_m(i,p)$ be the length of the shortest 
$\lambda$-pyramidal $(i,p)$-path which visits the cities $\overline{i,m}$. 
Due to the symmetry of the distance matrix $\Lambda_m(i,p)$ is also the length of a shortest 
$\lambda$-pyramidal $(p,i)$-path which visits the set of cities as above. 
In the special case $p=m$, $\Lambda_m(i,p)$ is the length of the monotone increasing path through the 
cities $\overline{i,m}$.

For $w,j,q\in \overline{1,n}$, $w\le j<q$, 
let $V_w(j,q)$ be the length of a shortest $\nu$-pyramidal $(j,q)$-path
visiting the cities $\overline{w,q}$ (and also the length of a shortest $\nu$-pyramidal $(q,j)$-path
visiting the above set of cities). 
In the special case $w=j$, $V_w(j,q)$ is the length of the monotone increasing path through the 
cities $\overline{j,q}$.

If $m=n$ or $w=1$ we omit the subscript $m$ or $w$ in $E_m(i,j)$, $\Lambda_m(i,p)$, $D_w(i,j)$, 
$V_w(j,q)$, and use simply $E(i,j)$, $\Lambda(i,p)$ and $D(i,j)$, $V(j,q)$, respectively.

Observe that for every $m\in \overline{2,n}$, the quantities $E_m(i,j)$ with $i,j \in \overline{1,m}$, $i<j$, 
can be computed in $O(m^2)$ time by the dynamic programming recursions (\ref{eq:dem1:m})-(\ref{eq:dem2:m}).
Analogously for every $w\in \overline{1,n-1}$, the quantities $D_w(i,j)$ with
$j,i\in \overline{w,n}$, $j<i$, can be computed in $O((n-w)^2)$ time by the recursions (\ref{eq:dem1:w})-(\ref{eq:dem2:w}). 

\begin{gather}
\label{eq:dem1:m}
E_m(i,j)=E_m(j,i)=\min
 \bigl\{ E_m(j+1,j)+c_{i,j+1},E_m(i,j+1)+c_{j+1,j} \bigr\},\\
\qquad \mbox{for $i\in \overline{1,m-2}$, $j\in \overline{i+1,m-1}$,} \mbox{ and } \nonumber \\
\label{eq:dem2:m}
E_m(i,m)=E_m(m,i):=c_{im}, \mbox{ for }
\qquad i\in \overline{1,m-1}.
\end{gather}
\begin{gather}
\label{eq:dem1:w}
D_w(i,j)=D_w(j,i)=\min
 \bigl\{ D_w(j-1,j)+c_{i,j-1},D_w(i,j-1)+c_{j-1,j} \bigr\},\\
\qquad \mbox{for $j\in \overline{w+1,n-1}$, $i\in \overline{j+1,n}$,} \mbox{ and } \nonumber \\
\label{eq:dem2:w}
D_w(i,w)=D_w(w,i):=c_{iw}, \mbox{ for }
\qquad i\in\overline{w+1,n}.
\end{gather}
In particular the entries $E(i,j)$ and the entries $D(j,i)$, $i,j\in\overline{1,n}$,
$i<j$, can be computed in $O(n^2)$ time.

Summarizing we get the following result:
\begin{observation}\label{Coro:Demi:m}
The quantities $E_m(i,j)$ for $i,j,m\in \overline{1,n}$, with $i<j\le m$, can be computed in $O(n^3)$ time. 
Analogously the quantities $D_w(i,j)$ for $i,j,w\in \overline{1,n}$, with $w\le j < i$, can be computed in $O(n^3)$ time. 
The quantities $E(i,j):=E_n(i,j)$ and $D(i,j):=D_1(j,i)$ with $i,j\in \overline{1,n}$, $i<j$ can be computed in $O(n^2)$ time.
\end{observation}

Since, as mentioned above, the optimal length of a pyramidal tour equals $E(1,2)+c_{21}$ 
Observation~\ref{Coro:Demi:m} implies the following result known already in the 1970's. 

\begin{theorem}(Klyaus~\cite{Klyaus}; Gilmore, Lawler \& Shmoys \cite{GLS})\label{theo:Demid}\\
A shortest pyramidal TSP tour visiting the cities $\overline{1,n}$ can be determined in $O(n^2)$ time. 
\end{theorem}

The quantities $\Lambda_m(i,p)$, with $i,p,m\in \overline{1,n}$, $ i<p\le m$, and $V_w(j,q)$, 
with $w,j,q\in \overline{1,n}$, $ w\le j<q$, can also be computed efficiently
by dynamic programming as shown in the following simple observation. 
We set $\Lambda_{m}(i,p):=0$ or $V_{w}(j,q):=0$, for $p=i=m$ or $w=j=q$, respectively. 

\begin{observation}
\label{pyramidalcomp:obse}
The quantities $\Lambda_m(i,p)$ with $i,p,m\in \overline{1,n}$, $ i<p\le m$, and $V_w(j,q)$ with 
$w,j,q\in \overline{1,n}$, $ w\le j<q$, can be computed in $O(n^3)$ time. 
In particular, for $m=n$ and $w=1$ the quantities $\Lambda(i,p)=\Lambda_n(i,p)$ with $i,p\in \overline{1,n}$, 
$i<p$, and $V(j,q)=V_1(j,q)$, with $j,q\in \overline{1,n}$, $ j<q$, can be computed in $O(n^2)$ time. 
\end{observation}
\proof
For the quantities $\Lambda_m(i,p)$ the claim follows directly from Observation~\ref{Coro:Demi:m} and 
the following equalities which hold for any triple $(i,p,m)$ in the given range of indices:
\[ \Lambda_m(i,p)=c_{i,i+1}+c_{i+1,i+2}+...+c_{p-2,p-1}+E_m(p-1,p)\, , \mbox{ if $i<p-1$, and }\]
\[ \Lambda_m(i,p)=E_m(p-1,p)\, , \mbox{ if $i=p-1$.}\]
Analogously, for the quantities $V_w(j,q)$ the claim follows directly from Observation~\ref{Coro:Demi:m} 
and the following equalities which hold for any triple $(w,j,q)$ in the corresponding range of indices:
\[ V_w(j,q)=c_{q,q-1}+c_{q-1,q-2}+...+c_{j+2,j+1}+D_w(j+1,j)\, , \mbox{ if $j<q-1$, and }\]
\[ V_w(j,q)=D_w(q,q-1)\, , \mbox{if $j=q-1$.}\]
Clearly the quantities $c_{i,i+1}+c_{i+1,i+2}+...+c_{p-2,p-1}$ can be computed in a preprocessing step 
in $O(n^3)$ time for all pairs $(i,p)$ with $i,p\in \overline{1,n}$ and $i<p$.
\qed

\smallskip
In what follows we assume that all quantities $E_m(i,j)$, $D_w(i,j)$,
$\Lambda_m(i,p)$, $V_w(j,q)$ with indices in their corresponding ranges are
computed in a preprocessing step. Moreover we use the straightforward relationships
$$E_m(i,j)=\min\Big\{c_{i,k}+\Lambda_m(k,j)\colon k\in \overline{j+1,m}\Big\}\ \ \mbox {for } i<j,\ \mbox {and} \ $$
$$ D_w(i,j)=\min\Big\{c_{i,k}+V_w(j,k)\colon k\in \overline{w,j-1}\Big\}\ \ \mbox {for } i>j. $$

\begin{figure}[bht]
\unitlength=1cm
\centering
\begin{picture}(15,6.9)
\put(0,1){\framebox[0.45\width]{
\centerline{\includegraphics[scale=0.9]{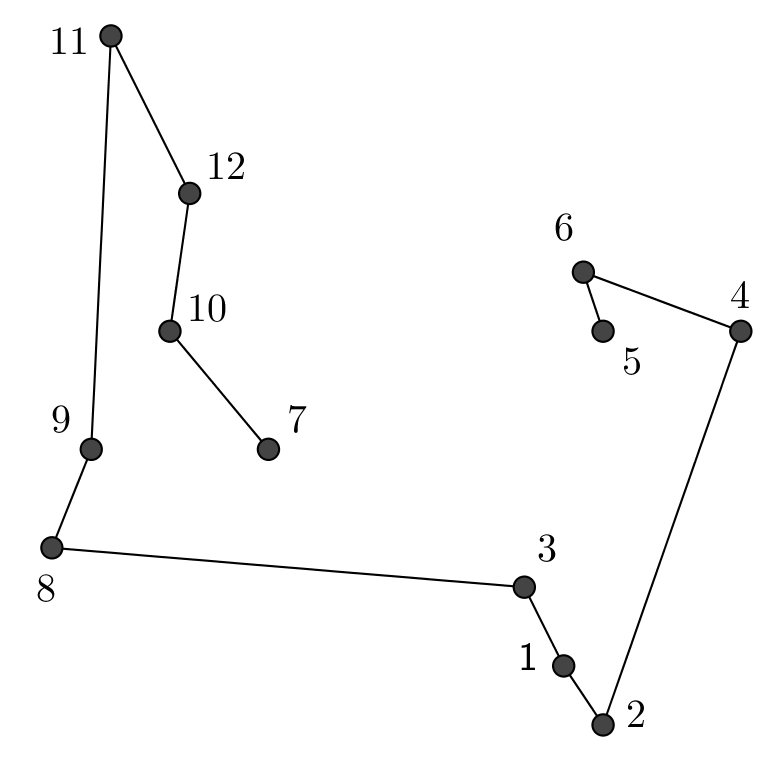}} }}
\put(4,1){
{ \centerline{\includegraphics[scale=0.9]{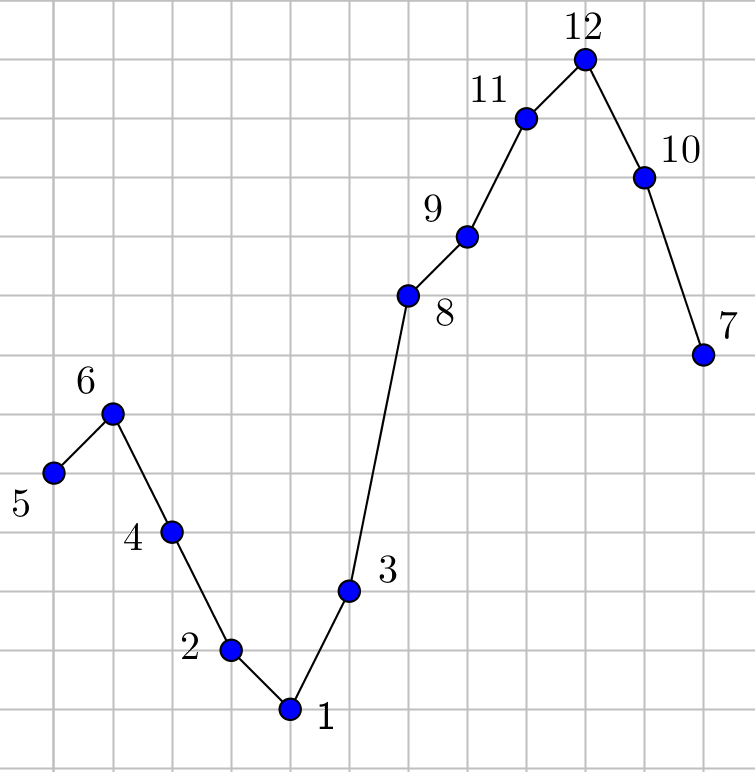}} }}
\put(3,0){(a)}
\put(12,0){(b)}
\end{picture}
\caption{
\label{fig:DemiPoints1}(a) The optimal (5,7)-TSP-path for the set of points in 
Figure~\ref{fig:DemiPpoints}; (b) Schematic representation of the path.}
\end{figure}

Finally let us notice that we will use a schematic representation of paths to
illustrate their combinatorial properties. 
An $(s,t)$-TSP-path $\tau=\seq{\tau_1=s,\ldots,\tau_k=t}$, $s<t$, visiting
the $n$ cities
$\tau_i\in \overline{1,n}$ for $i \in \overline{1,n}$, is visualized on an $n\times n$ grid by 
placing city $\tau_i$ in the grid node with coordinates $(i,\tau_i)$. 
For example, Figure~\ref{fig:DemiPoints1}(a) shows the shortest $(5,7)$-TSP-path $\tau$,
$\tau=\seq{5,6,4,2,1,3,8,9,11,12,10,7}$, for the set of $12$ points in Figure~\ref{fig:DemiPpoints}. 
The schematic representation of $\tau$ is shown in Figure~\ref{fig:DemiPoints1}(b).

\subsection{Classes of matrices}
\label{DefNot:matrices}
A symmetric $n\times n$ matrix $C=(c_{ij})$ is a \emph{Kalmanson matrix} if it satisfies the Kalmanson 
conditions in \eqref{eq:kalm1}--\eqref{eq:kalm2}, and  is a \emph{Demidenko matrix} if it satisfies 
the Demidenko conditions in \eqref{eq:demi}.
Note that a \emph{principal submatrix} of an $n\times n$ Kalmanson (Demidenko) matrix $C$ obtained from 
$C$ by deleting the rows and columns with indices in a subset $S\subset \overline{1,n}$, $|S|\le n-1$, 
is again a Kalmanson (Demidenko) matrix. 
The \emph{reversed matrix} $D=(d_{ij})$ results from matrix $C$ by simultaneously reversing the order 
of rows and columns in $C$; in other words, the matrix is specified by $d_{ij}=c_{n+1-i,n+1-j}$. 
A reversed Kalmanson (Demidenko) matrix is again a Kalmanson (Demidenko) matrix.

\begin{figure}[bht]
\unitlength=1cm
\centering
\begin{picture}(15,6.4)
\put(1,0.6){\framebox[0.5\width]{ \centerline{\includegraphics[scale=0.82]{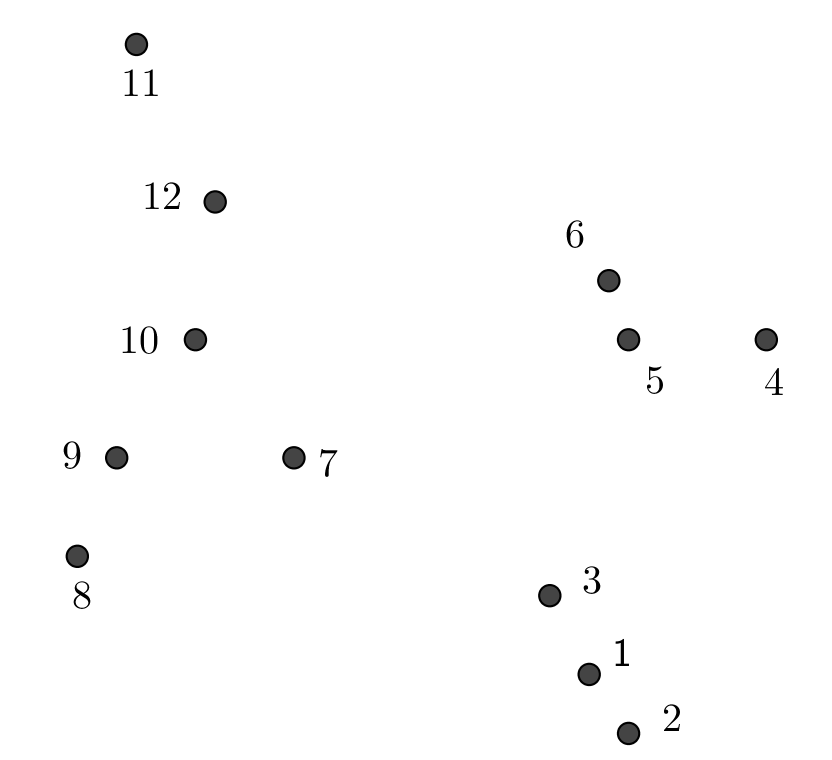}} }}
\small{
\put(10.5,3.1){\begin{tabular}{|r|r|} 
\hline
\# & (X,Y)\\
\hline \hline
1&(29,13)\\
2&(31,10)\\
3&(27,17)\\
4&(38,30)\\
5&(31,30)\\
6&(30,33)\\
7&(14,24)\\
8&(3,19)\\
9&(5,24)\\
10&(9,30)\\
11&(6,45)\\
12&(10,37)\\
\hline\hline 
\end{tabular}} 
}
\end{picture}
\caption{The distance matrix of this set of points is a Demidenko matrix but not a Kalmanson matrix.}
\label{fig:DemiPpoints}
\end{figure}

The example in Figure~\ref{fig:DemiPpoints} shows that the Demidenko matrices form proper superset of
the Kalmanson matrices. 
It can be checked that the distance matrix $C$ of the Euclidean distances of these $12$ points in the 
Euclidean plane is a Demidenko matrix but not a Kalmanson matrix. 
Indeed, some points lie far from the boundary of the convex hull of all points and the matrix of their 
Euclidean distances is not even a permuted Kalmanson matrix. 
We refer the reader to Deineko, Rudolf, Van der Veen \& Woeginger \cite{DRVW} for an explicit characterization 
of Euclidean sets of points that satisfy the Kalmanson conditions or the Demidenko conditions. 

Back in 1979, Demidenko~\cite{Demidenko1979} proved that an optimal TSP tour on a set of cities with a 
Demidenko distance matrix can be found among the pyramidal TSP-tours. 
Together with Theorem~\ref{theo:Demid} this implies the following result:
\begin{theorem}(Demidenko~\cite{Demidenko1979})\\
For an $n$-city TSP instance with a Demidenko distance matrix, an optimal TSP tour can be found
among the pyramidal TSP tours; hence it can be determined in $O(n^2)$ time. 
\end{theorem}

Throughout this paper, we will assume that all considered distance matrices have non-negative entries. 
This assumption can be made without loss of generality, as Demidenko  matrices can be transformed into 
non-negative Demidenko  matrices by simply adding a sufficiently large constant to each entry. 
Clearly the Path-TSP with a distance matrix $C=(c_{ij})$ and the Path-TSP with a distance matrix $\bar{C}=(c_{ij}+K)$ 
for some $K\in\rz$ are equivalent, in the sense that the sets of their optimal solutions coincide.
\section{Forbidden pairs of arcs and structural properties of optimal TSP-paths starting at city $1$}
\label{sec:structure} 
In this section we investigate the combinatorial structure of optimal
$(s,t)$-TSP-paths in the case where the distance matrix of the cities is a
Demidenko matrix.
An essential concept used in  our investigations is that of a \emph{forbidden
  pair of arcs}. We
show first that there always exists an optimal $(s,t)$-TSP-path which does not
contain 
forbidden pairs of arcs. As   implications of this  fact we obtain   the  solution of the
$(1,n)$-Path-TSP and further structural properties of the optimal
$(1,t)$-TSP-path for $t\neq 1$. 
 
\begin{definition} \label{forbidden:def}
In an $(s,t)$-path $\tau$ a pair of arcs $(i,\tau(i))$ and $(j,\tau(j))$ is called 
a \emph{forbidden pair of arcs} if either $i<j<\tau(i)<\tau(j)$ or $i>j>\tau(i)>\tau(j)$ holds.
\end{definition}

\begin{lemma}\label{forbid:Demid}
Consider a Path-TSP instance with a Demidenko distance matrix $(c_{ij})$.
There exists an optimal $(s,t)$-TSP-path which does not contain forbidden pairs of arcs.
\end{lemma}

\proof 
Let $\tau=\seq{\tau_1=s ,\tau_2,\ldots,\tau_n=t}$ be an optimal (i.e.\ shortest)
$(s,t)$-TSP-path with some forbidden pair
of arcs
$(i,\tau(i))$ and $(j,\tau(j))$. 
Assume without loss of generality  that $i<j<\tau(i)<\tau(j)$ and that city $i$ is reached
earlier than city $j$ in $\tau$.
Apply a standard transformation technique (see for instance Burkard \& al \cite{BSurv}) to construct
an optimal $(s,t)$-TSP-path which does not contain the pair $(i,\tau(i))$ and
$(j,\tau(j))$ of forbidden arcs: invert the $(\tau(i),j)$-subpath of $\tau$
into $\seq{j,\ldots,\tau(i)}$, and replace the forbidden pair of arcs by
the new pair of arcs $(i,j)$ and $(\tau(i),\tau(j))$. Clearly the resulting
path $\tau'$ is an $(s,t)$-TSP-path. Moreover the Demidenko conditions \eqref{eq:demi}
imply that
$c_{i\tau(i)}+c_{j\tau(j)}\ge c_{ij}+c_{\tau(i)\tau(j)}$, and therefore the
length of $\tau'$ does not exceed the length of $\tau$. So $\tau'$ is an optimal $(s,t)$-TSP-path
which does not contain the pair $(i,\tau(i))$ and $(j,\tau(j))$ of forbidden
arcs. If this path still contains a forbidden pair of arcs we apply the above
transformation again and repeat this process as long as the current optimal
$(s,t)$-TSP-path contains a forbidden pair of arcs. 

In order to see that this transformation process terminates after a final number of steps consider 
a potential function $K$ which maps any $(s,t)$-TSP-path $\pi$ to a non-negative integer 
$K(\pi):= \sum_{i=1,\ i\neq t}^n |i-\pi(i)|$. 
It can easily be seen that the transformation described above reduces the value of the potential
function, i.e.\ $K(\tau')<K(\tau)$. Since the potential function takes only non-negative integer 
values the process stops after a final number of steps.
\qed 

\smallskip
Lemma~\ref{forbid:Demid} implies the following result on the 
$(1,n)$-TSP-paths.
\begin{theorem} \label{Demi:1-n} 
 $\seq{1,2,\ldots,n}$ is a shortest $(1,n)$-TSP-path for the Path-TSP with a Demidenko distance matrix.
\end{theorem}
\proof
We show that any $(1,n)$-TSP-path with a non-trivial peak, i.e.\ a peak  different form $n$,  contains a forbidden pair of
 arcs. The proof of the lemma is then completed by observing that
$\seq{1,2,\ldots,n}$ is the unique $(1,n)$-TSP-path without a non-trivial peak. 

Let $m$ be the first peak in a $(1,n)$-TSP-path $\tau$. Since $m$ is a peak,
there is an arc $(i,j)$ in a 
$(m,n)$-subpath of $\tau$ such that $i<m<j$. If $\tau^{-1}(m)<i$, 
then $(\tau^{-1}(m),m)$ and $(i,j)$ build a forbidden pair of arcs. 
If $\tau^{-1}(m)>i$, then on the monotone increasing $(1,m)$-subpath of $\tau$ 
there exists an arc $(k,l)$ such that $k<i<l\le\tau^{-1}(m)$. By observing that
$\tau^{-1}(m)<m<j$ we conclude 
that the pair $(k,l)$ and $(i,j)$ is a forbidden pair of arcs in this case.
\qed
\begin{figure}[bht]
\unitlength=1cm
\centerline{\includegraphics[scale=0.9]{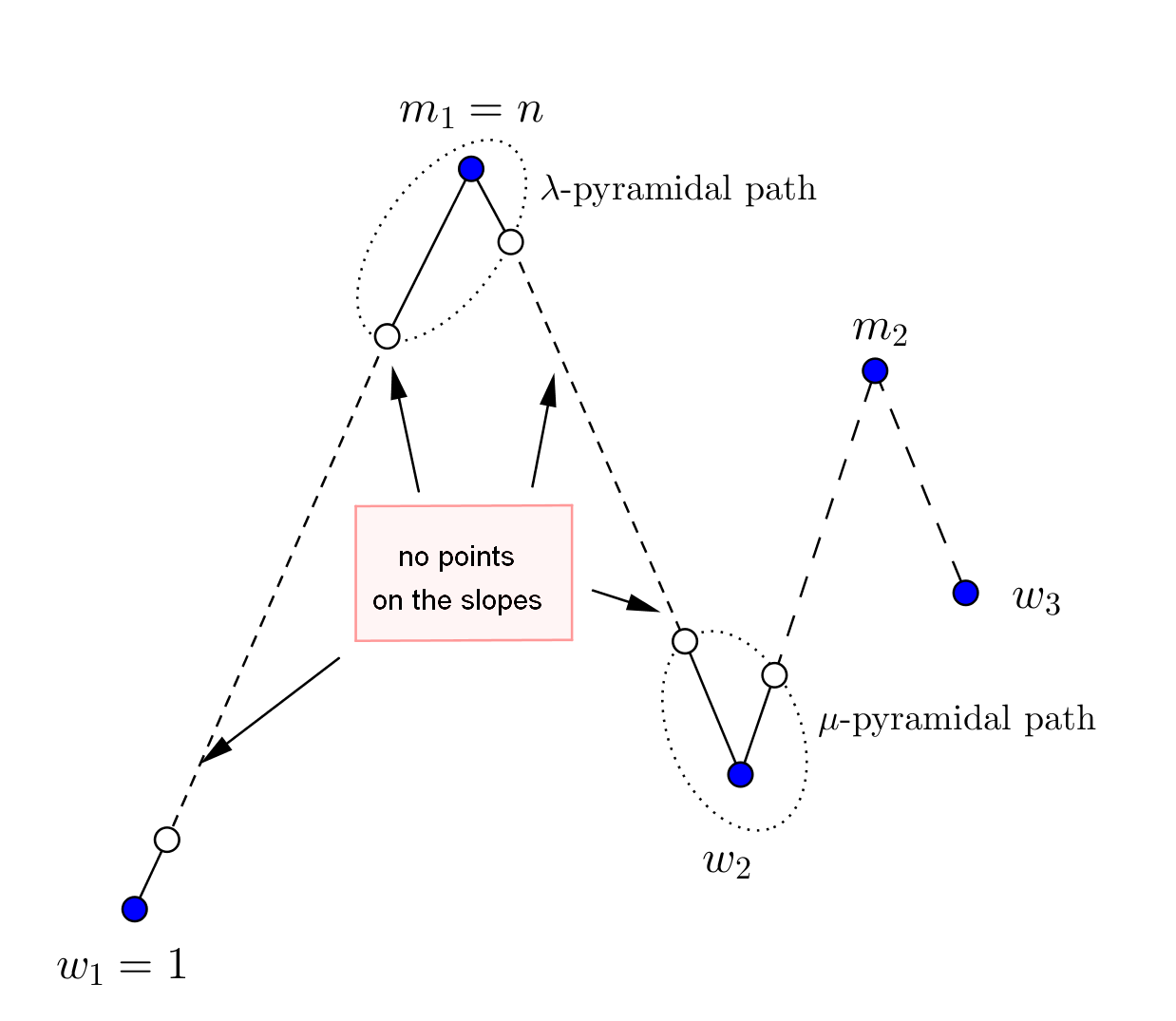}}
\caption{An illustration to the structure of a path without forbidden arcs.}
\label{fig:demi2}
\end{figure}

The following corollary is a statement about the monotonicity of peaks and
valleys in optimal $(1,t)$-TSP-paths. 
\begin{corollary}\label{Coro:Demid}
 Consider a Path-TSP on $n$ cities with a Demidenko distance matrix.
For any $t\in \overline{1,n}$, $t\neq 1$, there exists an optimal $(1,t)$-TSP-path with peaks
decreasing and valleys increasing from the left to the right in the path,
i.e.\ if peak $p$ (valley $v$) is reached earlier than peak $p'$ (valley $v'$)
in the path, than $p>p'$ ($v<v'$) holds.
\end{corollary}
\proof
The proof is done by induction on the number $n$ of cities. The statement is
trivially true for $n=2$. So assume that $n> 2$.

The correctness of the statement for $t=n$ follows immediately from
Theorem~\ref{Demi:1-n}: the $(1,n)$-TSP-path $\seq{1,2,\ldots,n}$ contains just
the (trivial) valley $1$ and the (trivial)  peak $n$.
Thus we assume without loss of generality that $t\neq n$ and let $\tau$ be an optimal
$(1,t)$-TSP-path. Let $\tau'=\seq{\tau_1=1,\tau_2,\ldots,\tau_k=n}$, with $k\in
\overline{1,n}$, $k<n$, be the $(1,n)$-subpath of $\tau$. Since a principal submatrix of
a Demidenko matrix is a Demidenko matrix,
as mentioned in Subsection~\ref{DefNot:matrices}, by applying Theorem~\ref{Demi:1-n} we can 
reorder the cities of $\tau'$ increasingly and obtain a $(1,n)$-path of the same length 
which visits the same cities as $\tau'$. 
So we can assume without loss of generality that there is no peak in the $(1,n)$-subpath $\tau'$ but
$n$, and hence $n$ is also the first peak in $(1,t)$-TSP-Path $\tau$. 
Now we distinguish two cases: (a) the last city $t$ is the smallest city in
the $(n,t)$-subpath of $\tau$ and (b) there is a city $j$ with $j<t$ in the
$(n,t)$-subpath of $\tau$. 
In Case (a) we can assume without loss of generality  that there are no other peaks but $n$ in the 
$(n,t)$-subpath of $\tau$ (by applying similar arguments to the one mentioned
above for the $(1,n)$-subpath $\tau'$ of $\tau$); this assumption is justified by Theorem~\ref{Demi:1-n} and by the fact 
that a reversed Demidenko matrix is a Demidenko matrix.
So. we assume w.l.o.og. that the $(n,t)$-subpath of $\tau$ is monotone
deareasing the statement of the corollary holds in this case. 
(Notice, that in this case we can find an optimal  $(1,t)$-TSP-path which  is $\lambda$-pyramidal.)

In Case (b) the $(n,t)$-subpath from $n$ to $t$ contains at least one valley
which is smaller than $t$. Let $v=\tau_l<t$ be the smallest valley in the
$(n,t)$-subpath of $\tau$, for $k<l<n$. 
Analogously as for the $(1,n)$-subpath and for Case (a) we can assume without loss of generality 
that the $(n,v)$-subpath of $\tau$ contains no peaks, but $n$. 
Denote by $\bar{\tau}$ the $(v,t)$-subpath of $\tau$, $1<v<t<n$. Since the
distance matrix of the cities visited by $\bar{\tau}$ is
a Demidenko matrix (as a principal submatrix of a Demidenko matrix), the
induction hypothesis applies and this completes the proof.
\qed

\begin{lemma}\label{lemma:slope} 
 Consider a Path-TSP with a Demidenko distance matrix and an optimal $(1,t)$-TSP-path
$\tau$ which contains no forbidden pairs of arcs and such that its peaks decrease and its valleys increase
 from the left to the right in the path. 
Let $m_1$ and $m_2$, $m_1>m_2$, be two consecutive peaks in $\tau$. Let $w_1$ be the valley that precedes peak $m_1$. 
let $w_2$ be the valley that follows $m_1$ and precedes $m_2$, and let $w_3$
be the valley that follows $m_2$.
Then the following statements hold:
\begin{itemize}
\item[(i)] The $(w_1,m_1)$-subpath of $\tau$ contains no
 city $i$ for which $w_2<i<m_2$ holds. 
\item[(ii)] The $(m_1,w_2)$-subpath of $\tau$ contains no city $j$ for which
 $w_3<j<m_2$ holds.
\end{itemize}
\end{lemma}

\proof 
Figure~\ref{fig:demi2} illustrates the structure of a path with the properties
described in the lemma. 
We prove here only statement (i), statement (ii) can be proved by using similar arguments.

Consider the $(w_1,m_1)$-subpath of $\tau$ and assume that (i) does not hold. 
 Let $i$ be the largest city on the $(w_1,m_1)$-subpath such that $w_2<i<m_2$. 
If $i>\tau^{-1}(m_2)$, then $\tau^{-1}(m_2)<i<m_2<\tau(i)$ and hence 
$(i,\tau(i))$, $(\tau^{-1}(m_2),m_2)$ build a forbidden pair of arcs
contradicting the assumption of the lemma. If $i<\tau^{-1}(m_2)$, then on the
$(w_2,\tau^{-1}(m_2))$-subpath of $\tau$ 
there exists an arc $(k,l)$ such
that $k<i<l$. In this case $(i,\tau(i))$ and $(k,l)$ build a forbidden
pair of arcs and this contradicts the assumption of the lemma. 
\qed

\smallskip
In particular, the statements in the above lemma imply  that the cities $\overline{m_2+1,m_1}$ 
are placed on consecutive positions in path $\tau$ and form a $\lambda$-pyramidal subpath of it.

\section{Efficient solution of the Path-TSP with a Demidenko distance matrix: the case $s=1$ }
\label{sec:s=1}

In this  section we first  derive recursive equations for the length $H_n(1,t)$ of an optimal 
$(1,t)$-TSP-path through the cities $\overline{1,n}$. These equations yield an $O(n^2)$
dynamic programming algorithm  for the solution of the Path-TSP with a
Demidenko distance matrix where the   starting city
is $1$ and 
the  destination city arbitrary.

{\bf The recursive equations for $H_n(1,t)$.} Due to Lemma~\ref{forbid:Demid}, Corollary~\ref{Coro:Demid} and Lemma~\ref{lemma:slope}, we consider 
without loss of generality an optimal $(1,t)$-TSP-paths $\tau$ which contains no forbidden pairs of arcs, 
has decreasing peaks and increasing valleys from the left to the right and fulfills the statements of 
Lemma~\ref{lemma:slope}.
We distinguish two cases: 
(1) the optimal $(1,t)$-TSP-path contains no valleys but $1$ and $t$, and 
(2) the optimal $(1,t)$-TSP-path contains at least one valley $w$ with $1<w<t$.
In the first case the optimal $(1,t)$-TSP-path is $\lambda$-pyramidal and thus $H_n(1,t)=\Lambda_n(1,t)$. 
In the second case let $w= j+1$, $j\ge 1$, $w<t$ be the left-most (and the
smallest) non-trivial valley in $\tau$.

Since $n$ is the first peak, valley $w$ is reached after $n$ in $\tau$.
According to Lemma~\ref{lemma:slope} (see also Figure~\ref{fig:demi2} where the
role of $w$ is played by $w_2$), $\tau$ starts with a monotone increasing 
$(1,w-1)$-subpath visiting the cities $\overline{1,j}$, 
followed first by a $\lambda$-pyramidal path with the peak $m=n$, then by a $\mu$-pyramidal path with valley 
$w=j+1$, and so on, until the final city $t$ is reached. 
For cities $j$, $m$, $w$ such that $j<w<m$ and $w\le t<m$ denote by $\Gamma(j,m,w)$ be the length of an optimal
$(j,t)$-path starting at $j$ and then visiting the cities of 
the set $\{j\}\cup\overline{w,m}$ with the first peak in this path being
$m$, and the valley that follows $m$ being $w>j$. Here $w$ could
also coincide with $t$, in which case $t$ would be reached along a decreasing
sequence of cities from $m$ to $t$. Then the length $H_n(1,t)$ of the optimal
$(1,t)$-TSP-path is given as follows

\begin{equation}
\label{eq:H}
\begin{aligned}
H_n(1,t)=\min \big \{ \Lambda_n(1,t), \min\{V_1(1,j)+\Gamma(j,n,j+1)\colon j=1,\ldots,t-2\}
 \big \} \, . 
\end{aligned}
\end{equation}
where $V_1(1,j)$ is defined as in Subsection~\ref{DefNot:path}.

Next we give a  recursive equation for the computation of the  quantities
$\Gamma$ above. To this end denote by $L(k,w,p)$  the length of an optimal 
$(k,t)$-path visiting the cities of
 the set $\overline{w,p}\cup\{k\}$ with first valley $w$ and first peak
 $p$ such that $w$ precedes $p$, for cities $w$, $p$, $k$ such that $w<p<k$ and
 $w<t\le p$.
Here $p$ could
also coincide with $t$ in which case $t$ would be reached along an increasing
sequence of cities from $w$ to $t$.

\begin{figure}[bht]
\unitlength=1cm
\begin{picture}(15,10)
\put(-3,6)
{\centerline{\includegraphics[scale=1]{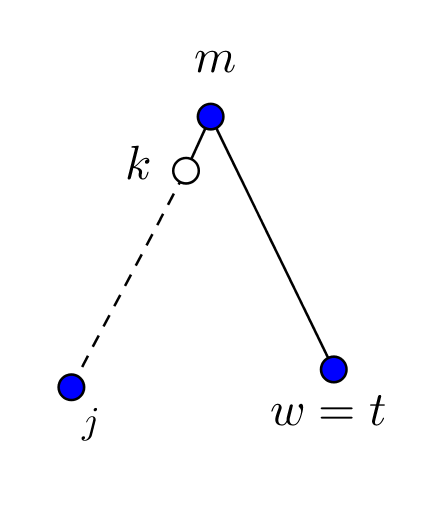}}}
\put(0,6){\mbox{$\Gamma(j,m,w)=E_m(j,t)=min_{k\in \overline{t+1,m}}\{c_{j,k}+\Lambda_m(k,t)\}$}}
\put(3.8,5.5){\mbox{(a)}}
\put(4.5,6)
{\centerline{\includegraphics[scale=1]{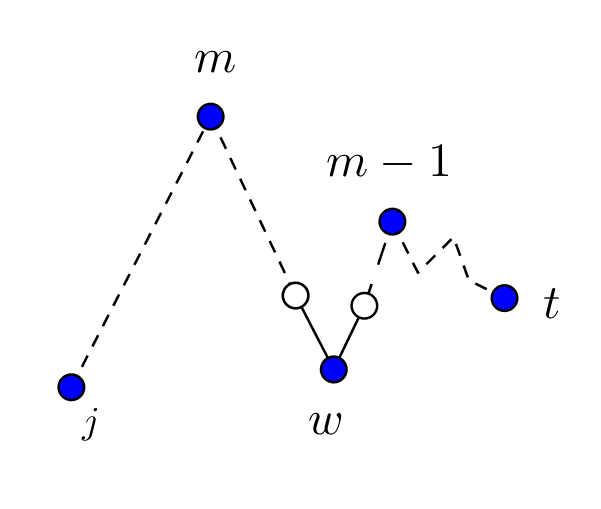}}}
\put(9.8,6){\mbox{$\Gamma(j,m,w)=c_{j,m}+L(m,w,m-1)$}}
\put(11.3,5.5){\mbox{(b)}}
\put(-3,1)
{\centerline{\includegraphics[scale=1]{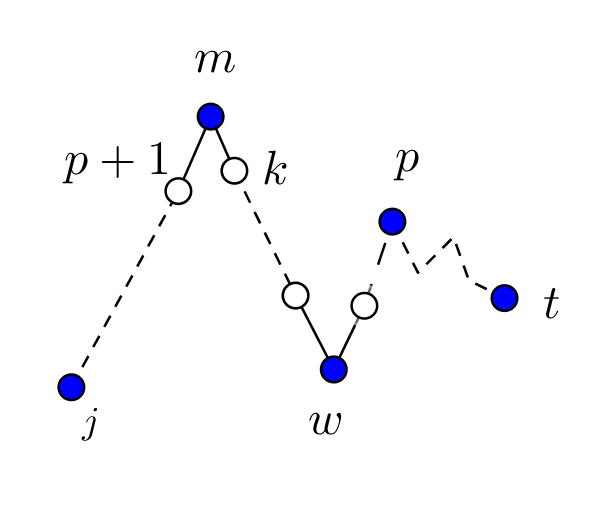}}}
\put(0,0.6)
{\mbox{$\Gamma(j,m,w)=c_{j,p+1}+\Lambda_m(k,p+1)+L(k,w,p)$}}
\put(3.4,0){\mbox{(c)}}
\put(4.2,1){\centerline{\includegraphics[scale=1]{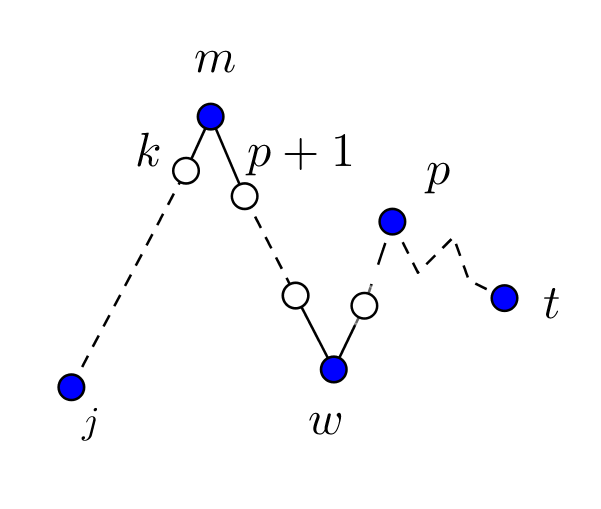}}}
\put(8.5,0.6)
{\mbox{$\Gamma(j,m,w)=E_m(j,p+1)+L(p+1,w,p)$}}
\put(11.3,0){\mbox{(d)}}
\end{picture}
\caption{An illustration to the calculation of $\Gamma(j,m,w)$. The straight dashed lines represent single edges.}
\label{fig:demi3}
\end{figure}
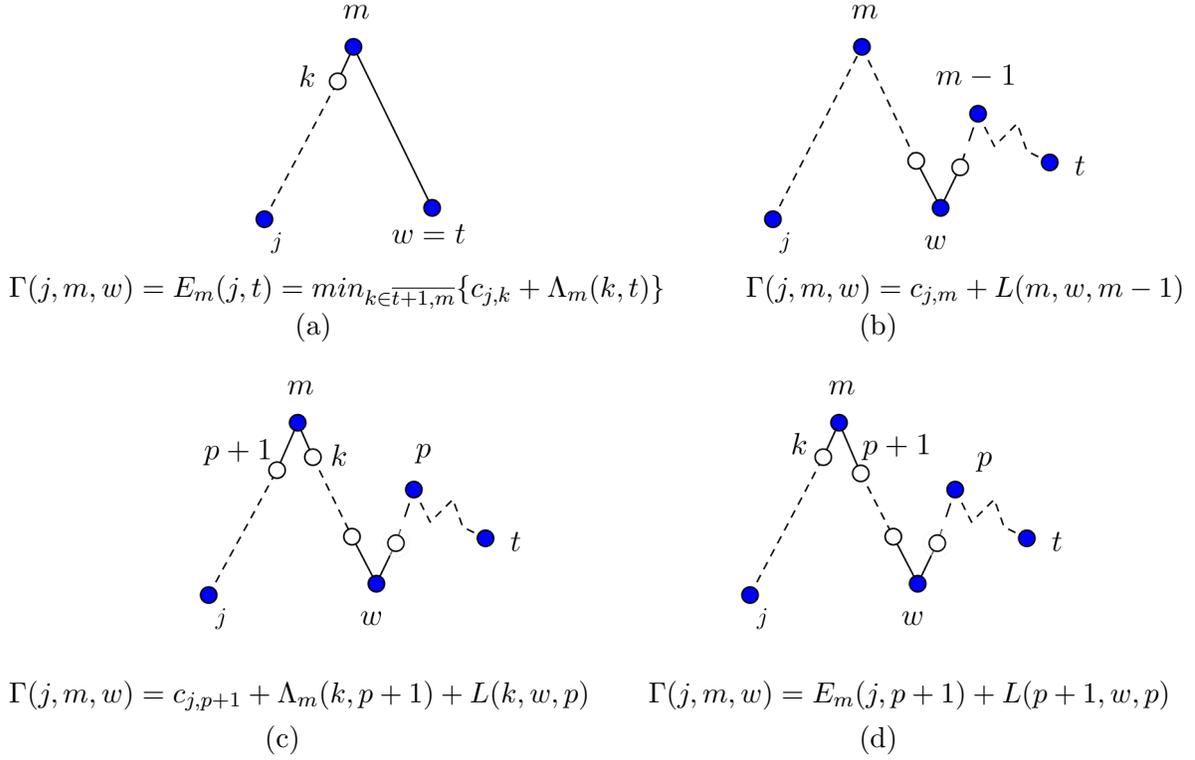

According to Lemma~\ref{forbid:Demid}, Corollary~\ref{Coro:Demid} and
Lemma~\ref{lemma:slope} and as illustrated in Figure~\ref{fig:demi3}
the values $\Gamma(j,m,w)$, for $j,m,w\in \overline{1,n}$ with $ j<w\le t <m$, can be calculated by the following recursions:
 
\begin{align}
\nonumber
&\Gamma(j,m,w) = 
E_m(j,t)\, , \mbox{ if $w=t$, and } \\
\label{eq:Gamma}
&\Gamma(j,m,w) = 
 \min
 \begin{cases}
 c_{j,m}+L(m,w,m-1), 
\\
 \min\big \{\alpha_{j,m,w}(p,k)\colon p\in\overline{t,m-2},
 k\in\overline{p+2,m}\big \}
 \end{cases}, \mbox{ if $w<t$},
\end{align}
where $\alpha_{j,m,w}(p,k):=\min\{ c_{j,p+1}+\Lambda_m(k,p+1)+L(k,w,p), 
 E_m(j,p+1)+L(p+1,w,p)\}$.

\smallskip
The above recursions (\ref{eq:Gamma}) cover the following cases
which are also  illustrated in Figure~\ref{fig:demi3}. 

\smallskip
Case $w=t$. The
$(j,w)$-path has only one peak $m$ (since $w=t$), starts at $j$, ends at $t$
and goes through the cities
 $\overline{t+1,m}$. The optimal length of the pyramidal path on these
 cities equals $E_m(j,t)$ (see the definition before Observation~\ref{Coro:Demi:m}). 
This case corresponds to the first line in (\ref{eq:Gamma}) and is illustrated in Figure~\ref{fig:demi3}(a). 

\smallskip
Case $w\neq t$. In this case there will be another peak (different from $m$) to the right of
the valley $w$. 
Let this peak be $p$, with $p\le m-1$ according to
Corollary~\ref{Coro:Demid}. We again distinguish two subcases: $p=m-1$ and
$p<m-1$. 
If $p=m-1$, then according to Lemma~\ref{lemma:slope} the path starts 
with the arc $(j,m)$ followed by $(m,t)$-subpath
which starts at $m$ and goes through the cities $\overline{w,m-1}$, with
the first valley in this path being $w$ and peak $m-1$ reached after valley
$w$.
 The optimal length of such a path is given by $L(m,w,m-1)$ and in this case $\Gamma(j,m,w)$ is calculated as shown in 
the second line of (\ref{eq:Gamma}) and illustrated in
Figure~\ref{fig:demi3}(b).
\smallskip

 If $p<m-1$, then the $(j,t)$-path 
starts with an arc connecting $j$ to the
left-most city of a $\lambda$-pyramidal path with peak $m$ containing the cities 
 $\overline{p+1,m}$ and either starting or ending at city $p+1$. 
Let the other end-city of this $\lambda$-pyramidal path be $k\in
\overline{p+2,m}$. 
Then the length of the $(j,t)$-path is calculated as shown in the third line of (\ref{eq:Gamma}) and 
illustrated in Figures~\ref{fig:demi3}(c) and~\ref{fig:demi3}(d). 
These pictures correspond to the cases where above mentioned $\lambda$-pyramidal path starts 
or ends at $p+1$, respectively. 
 
\smallskip
The values $L(k,w,p)$, for $w,k,p\in \overline{1,n}$ with $1<w< t \le p< k\le n$, 
can be computed recursively in a similar way:

\hspace*{-2cm}
\begin{align}
\nonumber
&L(k,w,p)=D_w(k,t)\, , \mbox{ if $p=t$, and}\\
\label{eq:L}
&L(k,w,p)=
 \min \begin{cases}
\min \big \{ \beta_{k,w,p}(v,j)
 \colon v\in\overline{w+2,t}, 
 j\in\overline{w,v-2}\big \}\\
 c_{k,w}+\Gamma(w,p,w+1)
 \end{cases}, \mbox{if $p>t$,}
\end{align}
where
$\beta_{k,w,p}(v,j):=\min\{D_w(k,v-1)+\Gamma(v-1,p,v),c_{k,v-1}+V_w(j,v-1)+\Gamma(j,p,v)\}$.

\smallskip
Based on equations~(\ref{eq:H})-(\ref{eq:L}) we obtain the following result about the computation of the optimal $(1,t)$-TSP path
in the case of a Demidenko distance matrix.
\begin{theorem}\label{maintheo:Demid}
 Consider a Path-TSP on $n$ cities $\overline{1,n}$ with a Demidenko distance matrix.
An optimal $(1,t)$-TSP-path can be found in $O(n^5)$ time.
\end{theorem}
\proof 
Recall that all values $\Lambda_m(i,p)$, for $i,p,m\in \overline{1,n}$ with $i<p\le m$,
 and $V_w(j,q)$, for $w,j,q\in \overline{1,n}$, with $w\le j<q$, can be calculated
in $O(n^3)$ time in a preprocessing step, see Observation~\ref{pyramidalcomp:obse}.
Further, according to equation (\ref{eq:H}) the computation of the length $H_n(1,t)$ of the optimal
$(1,t)$-TSP-path (for $t\ge 3$) involves the quantities
$\Gamma(j,n,j+1)$, for $j\in \overline{1,t-2}$. The quantities
$\Gamma(j,m,w)$, for $j,w,m\in \overline{1,n}$ with $j<w\le t <m$, are computed
recursively together with the quantities $L(k,w,p)$, for $w,p,k\in \overline{2,n}$ with $w< t \le p< k$. 
Observe that the computation of $\Gamma(j,m,w)$, for some $(j,m,v)$ in the corresponding range,
just involves quantities $L(x,y,z)$ for which the difference $z-y$ between the specified peak $z$ 
and valley $y$ is strictly smaller than the difference $m-w$ between the specified peak $m$ and 
the specified valley $w$ in $\Gamma(j,m,w)$.
Analogously, the computation of $L(k,w,p)$, for some $(k,w,p)$ in the corresponding range, involves quantities
$\Gamma(x,y,z)$ for which the difference $y-z$ between the specified peak $y$ and valley $z$ is strictly 
smaller than the difference $p-w$ between the specified peak $p$ and the specified valley $w$ in $L(k,w,p)$. 
So the recursion would start with the trivial values $\Gamma(j,t+1,t):=c_{j,t+1}+c_{t+1,t}$ for $j,t\in \overline{1,n}$ 
with $ j<t$, and $L(k,t-1,t):=c_{k,t-1}+c_{t-1,t}$ for $k,t\in \overline{1,n}$ with $k>t$. 
The effort needed for the recursive computation of all other values of $\Gamma$ and $L$ is the computation of the 
minimum over $O(n^2)$ sums consisting of previously computed entries of the same arrays $L$, $\Gamma$ and of 
appropriate entries of the arrays $E$, $\Lambda$, $D$, $V$ all of which have been computed in a preprocessing step. 
Hence the computation of each of the $O(n^3)$ values $\Gamma$ ($L$) can be done in $O(n^2)$ time 
and the overall time complexity is $O(n^5)$.
\qed

\section{Efficient solution of the Path-TSP with a Demidenko distance matrix: the general case }
\label{sec:s=general}

In this section deals with  the  polynomial time computation of an optimal $(s,t)$-TSP-path in 
the most general case where $s,t\in\overline{1,n}$, and none of the conditions $s=1$ and 
$t=n$ is necessarily fulfilled. The results is stated in Theorem~\ref{theo:Demid_s_t}
and the rest of the section is dedicated to  its proof. 

The following assumprions hold throughout the rest of this section.
Due to the symmetry of the distance matrix we can assume  that  $s<t$ holds.
Moreover, we assume  that $s>1$ because the case $s=1$ has already been handled in Theorem~\ref{maintheo:Demid}. 
The case $t=n$ can be handled analogously to the case $s=1$, since  a reversed Demidenko matrix is
again a Demidenko matrix, as  pointed out in Section~\ref{sec:DefNot}. 
Summerizing we  assume without loss of generality that $1<s<t<n$ holds. 
Further, by Lemma~\ref{forbid:Demid} we only consider   $(s,t)$-TSP-paths without forbidden pairs of arcs.
The following three claims state  some particular structural
properties of such paths which will be useful for the proof of  Theorem~\ref{maintheo:Demid}.
\smallskip

\begin{claim}\label{claim1}
There is an optimal $(s,t)$-TSP-path $\tau$ with $1<s<t<n$, where city $1$ precedes city $n$. 
\end{claim}
\proof
To prove the claim we show that any $(s,t)$-TSP-path $\tau$ in which $n$ precedes $1$ contains a forbidden pair of arcs. 
Indeed, the  $(s,n)$-subpath of $\tau$ contains an arc $(u,v)$ with $u<p<v$, where $p$ is
the first peak in the subpath from $1$ to $t$. (If the subpath from $1$ to $t$ is monotone we have $p=t$.)
If $\tau^{-1}(p)<u$, then the arcs $(u,v)$ and ($\tau^{-1}(p),p)$ build a
forbidden pair of arcs. Otherwise consider an arc $(x,y)$ in the (monotone)
subpath from $1$ to $p$ such that $x<u<y$. 
The arcs $(u,v)$, $(x,y)$ build a forbidden pair of arcs in this case.
\qed

\begin{claim}\label{claim2} There is an optimal $(s,t)$-TSP-path in which $1$ precedes $n$ and each city 
  of the $(s,1)$-subpath is smaller than each city of the $(n,t)$-subpath.
  \end{claim}
\proof
To prove the claim we consider an arbitrary $(s,t)$-TSP-path $\tau$ in which city $1$ precedes city $n$ 
and show that the existence of a city in the $(s,1)$-subpath which is larger than some city in the 
$(n,t)$-subpath implies the existence of a forbidden pair of arcs.
In particular if $i$ is a city in the $(s,1)$-subpath of $\tau$ and $j$ is a city in the 
$(n,t)$-subpath of $\tau$ such that $i>j$, it can be shown by arguments similar to those in 
the proof of Claim~\ref{claim1} that the $(i,j)$-subpath of $\tau$ contains a forbidden pair of arcs.
\qed

\begin{claim}\label{claim3}
  There exists an optimal $(s,t)$-TSP-path $\tau$ which is a concatenation
of two paths $\tau_1^{p}$ and $\tau_2^p$ such that  $\tau_1^p$ starts at $s$ and visits
all cities from the set $\overline{1,p-1}$, and  $\tau_2^p$ starts at the last city of $\tau_1^p$, 
then visits all cities from the set $\overline{p,n}$ and ends at $t$, for some $p\in \overline{s+1,t}$. 
\end{claim}
\proof
To prove the claim we consider an $(s,t)$-TSP-path $\tau$ in which city $1$ precedes city $n$ and such 
that each city of the $(s,1)$-subpath of $\tau$ is smaller than each city of the $(n,t)$-subpath of $\tau$. 
Claim~\ref{claim2} guarantees the existence of such a path.
Observe that due to Theorem~\ref{Demi:1-n} and the fact that a principal submatrix
of a Demidenko matrix is a Demidenko matrix, the $(1,n)$-subpath of $\tau$ is monotone increasing. 
Now let $p$ be the smallest city on the $(n,t)$-subpath of $\tau$ and let $x$ be the last city 
with $x<p$, $x\neq s$, in the $(1,n)$-subpath of $\tau$. 
Then the cities in $\overline{1,p-1}$ are exactly the cities visited by the $(s,x)$-subpath of $\tau$. 
Hence we can set $\tau_1^p$ and $\tau_2^p$ as the $(s,x)$-subpath and the $(x,t)$-subpath of $\tau$, respectively.
\qed
\begin{theorem}
\label{theo:Demid_s_t}
Consider a Path-TSP on $n$ cities $\overline{1,n}$ with a Demidenko distance matrix
and a given pair of cities $(s,t)$.
An optimal $(s,t)$-TSP-path can be found in $O(|t-s|n^5)$ time.
\end{theorem}
\proof
Due to Claims~\ref{claim1}-\ref{claim3} we minimize over $(s,t)$-TSP-paths
$\tau$  in which city $1$ precedes city $n$, each city in the $(s,1)$-subpath has a
smaller index than each city in the  $(n,t)$-subpath, and for which an index
$p\in \overline{s+1,t}$ exists such that $\tau$ is the concatenation of
$\tau_1^p$ and $\tau_2^p$ as described in Claim~\ref{claim3}. We refer to $\tau_1^p$ and $\tau_2^p$
as the prefix and the postfix of $\tau$ and denote by $x$ the last city of
the prefix (and the first city of the postfix).  Next we  show how to efficiently determine the length of a shortest 
prefix and a shortest postfix for each $p\in \overline{s+1,t}$ and each $x$, $x<p$, $x\ne s$. 
 
By using the fact that a principal submatrix of a Demidenko matrix is a Demidenko matrix and by 
considering that the postfix starts at $x$, $x<p$, and then visits all cities in $\{x\}\cup\overline{p,n}$, 
the shortest length $T(x,p)$ of the postfix can be determined in $O(n^5)$ as described in Theorem~\ref{maintheo:Demid},
for every $p\in\overline{s+1,t} $ and for every  $x<p$.
Now let us virtually shrink the postfix to its second city $\tau(x)$. 
To distinguish $\tau(x)$ from the shrunk city let us denote the later by ${\mathcal V}_x^p$. 
We set the distance between ${\mathcal V}_x^p$ and $x$ equal to $T(x,p)$. 
Observe that the index of city  $\tau(x)$ is larger than the indices of all
cities in the prefix (with indices lying in $\overline{1,p-1}$), 
hence the dummy city ${\mathcal V}_x^p$  can be considered to have index   $p$.
Consider now the path $\bar{\tau}_1^p$ obtained by extending the prefix along the edge $(x,{\mathcal V}_x^p)$;
since $\tau$ contains no forbidden pairs of arcs also $\bar{\tau}_1^p$ contains no forbidden pairs of arcs.
Notice that a shortest $(t,n)$-TSP-path without a forbidden pair of arcs can be determined in the same
way as a shortest $(1,t)$-TSP-path without a forbidden pair of arcs (after renumbering the rows and 
columns of the distance matrix from the right to the left prior to the calculations). 
Thus the length $\bar{T}(x,p)$ of $\bar{\tau}_1^p$ which starts at $s$, visits all cities from 
$\overline{1,p-1}$ and then ends at ${\mathcal V}_x^p$ which has index $p$,  can be computed in $O(n^5)$ time as described 
in Theorem~\ref{maintheo:Demid} (recall that  the proof of Theorem~\ref{maintheo:Demid} relies exclusively on 
the properties of paths without forbidden pairs of arcs). 
Clearly, $\bar{T}(x,p)$ equals the sum of the lengths of the prefix and the postfix, 
given $p\in\overline{s+1,t} $ and $x<p$, $x\ne s$.
Consequently, the required length of the shortest $(s,t)$-TSP-path equals 
$\min\{ \bar{T}(x,p): p\in\overline{s+1,t}\, , x<p\, , x\neq s\}$. 
It is straightforward to compute this minimum in $O((t-s)n^6)$ time after having computed 
$\bar{T}(x,p)$ in $O(n^5)$ time for each pair of indices $x$ and $p$ as above. 
A closer look at the recursions (\ref{eq:H}) and equations (\ref{eq:Gamma})-(\ref{eq:L}) involved 
in the computation of $\bar{T}(x,p)$ and $T(x,p)$ according to Theorem~\ref{maintheo:Demid}, reveals 
that for each fixed $p$ all computations use the same quantities $\Gamma$ and $L$, independently 
on the value of $x$, where $x<p$ and $x\neq s$. 
Thus for any $p
\in\overline{s+1,t}$ all values of $\bar{T}(x,p)$ (and also $T(x,p)$) for $x<p$, $x\neq s$, 
can be computed in $O(n^5)$ time leading  to an overall time complexity of
$O((t-s)n^5)$ and  completing  the proof.
\qed

\section{Final remarks}
\label{sec:remarks}
We have analyzed the Path-TSP on Demidenko matrices, and we have derived a sophisticated
algorithm with a polynomial time complexity of roughly $O(n^6)$.
An obvious open problem is to get an improvement to some more civilized time complexity 
like $O(n^2)$, or at least $O(n^3)$.

The polynomially solvable special case for Demidenko matrices can be used in local search 
approaches for the Path-TSP, as it yields an exponential neighborhood over which we can 
optimize in polynomial time. 
We refer to
Ahuja, Ergun, Orlin \& Punnen \cite{Ahuja},
Deineko \& Woeginger \cite{Neighborhoods},
Gutin, Yeo \& Zverovich \cite{GutinYeo}, and
Orlin \& Sharma \cite{Orlin}
for a discussion of similar approaches in the case of the classical TSP.

\medskip
\paragraph{Acknowledgements.}
This research was conducted while Vladimir Deineko and Gerhard Woeginger were visiting TU Graz,
and they both thank the Austrian Science Fund (FWF): W1230, Doctoral Program in ``Discrete Mathematics'' 
for the financial support.
Vladimir Deineko acknowledges support 
by Warwick University's Centre for Discrete Mathematics and Its Applications (DIMAP).
Gerhard Woeginger acknowledges support
by the DFG RTG 2236 ``UnRAVeL'' -- Uncertainty and Randomness in Algorithms, Verification and Logic.



\medskip
\begin{thebibliography}{28}

\bibitem{Ahuja}
{\sc R.K. Ahuja, O. Ergun, J.B. Orlin, and A.P. Punnen} (2002). 
A survey of very large-scale neighborhood search techniques.
\emph{Discrete Applied Mathematics 123}, 75–102.

\bibitem{Applegate} 
{\sc D.L. Applegate, R.E. Bixby, V. Chv\'{a}tal, and W.J. Cook} (2006).
\emph{The Traveling Salesman Problem: A Computational Study}, 
Princeton University Press.

\bibitem{BSurv}
{\sc R.E. Burkard, V.G. Deineko, R. van Dal, J.A.A. van der Veen, and G.J. Woeginger} (1998).
Well-solvable special cases of the TSP: a survey.
\emph{SIAM Reviews 40}, 496--546.

\bibitem{Cook} 
{\sc W.J. Cook} (2012).
\emph{In Pursuit of the Travelling Salesman: Mathematics and the Limits of Computation}. 
Princeton University Press. 

\bibitem{DeKlTiWo2014}
{\sc V.G. Deineko, B. Klinz, A. Tiskin, and G.J. Woeginger} (2014).
Four-point conditions for the TSP: the complete classification.
\emph{Discrete Optimization 14}, 147--159.

\bibitem{DRVW}
{\sc V.G. Deineko, R. Rudolf, J.A.A. Van der Veen, and G.J. Woeginger} (1997).
Three easy special cases of the Euclidean traveling salesman problem.
\emph{RAIRO Operations Research 31(4)}, 342--362.

\bibitem{Neighborhoods}
{\sc V.G. Deineko and G.J. Woeginger} (2000).
A study of exponential neighborhoods for the travelling salesman problem and for the quadratic assignment problem. 
\emph{Mathematical Programming 87}, 519--542.

\bibitem{Demidenko1979}
{\sc V.M. Demidenko} (1979).
The travelling salesman problem with asymmetric matrices.
\emph{Vestsi Akad.\ Navuk BSSR Ser.\ Fiz.-Mat.\ Navuk 1}, 29---35, (in Russian).

\bibitem{Garcia1}
{\sc A. Garcia and J. Tejel} (1996).
Using total monotonicity for two optimization problems on the plane.
\emph{Information Processing Letters 60}, 13--17.

\bibitem{Garcia2}
{\sc A. Garcia, P. Jodra, and J. Tejel} (1998).
An efficient algorithm for on-line searching in Monge path-decomposable tridimensional arrays.
\emph{Information Processing Letters 68}, 3--9.

\bibitem{GJ}
{\sc M.R. Garey and D.S. Johnson} (1979).
\emph{Computers and Intractability: A Guide to the Theory of NP-Completeness}.
Freeman, San Francisco.

\bibitem{GLS}
{\sc P.C. Gilmore, E.L. Lawler, and D.B. Shmoys} (1985).
Well-solved special cases.
Chapter~{4} in \emph{The Traveling Salesman Problem}, E.L. Lawler, J.K. Lenstra,
A.H.G. Rinnooy Kan, and D.B. Shmoys (eds.), Wiley, Chichester, 87--143.

\bibitem{GutinYeo} 
{\sc G. Gutin, A. Yeo, and A. Zverovich} (2002). 
Exponential neighborhoods and domination analysis for the TSP. 
Chapter 6 in \emph{The travelling salesman problem and its variations}, 
G. Gutin and A.P. Punnen (eds.), Kluwer Academic Publishers, 223--256. 

\bibitem{Gutin} 
{\sc G. Gutin and A.P. Punnen} (2002).
\emph{The travelling salesman problem and its variations}. 
Kluwer Academic Publishers.

\bibitem{Kabadi} 
{\sc S.N. Kabadi} (2002).
Polynomially solvable cases of the TSP. 
Chapter~11 in \emph{The travelling salesman problem and its variations}, 
G. Gutin and A.P. Punnen (eds.), Kluwer Academic Publishers, 489--583.

\bibitem{Kalmanson1975}
{\sc K.\ Kalmanson} (1975).
Edgeconvex circuits and the travelling salesman problem,
\emph{Canadian Journal of Mathematics 27}, 1000--1010.

\bibitem{Klyaus}
{\sc P.S. Klyaus} (1976). 
The structure of the optimal solutions of some classes of the traveling salesman problem.
\emph{Izv.\ Akad.\ Nauk.\ BSSR, Ser.\ Fiz.-mat.\ Nauk 6}, 95--98, (in Russian).

\bibitem{TSP-book}
{\sc E.L. Lawler, J.K. Lenstra, A.H.G. Rinnooy Kan, and D.B. Shmoys} (1985).
\emph{The Traveling Salesman Problem}.
Wiley, Chichester.

\bibitem{Orlin} 
{\sc J.B. Orlin and D. Sharma} (2004).
The extended neighborhood: Definition and characterization. 
\emph{Mathematical Programming 101}, 537--559.

\bibitem{PaYa93}
{\sc C.H. Papadimitriou and M. Yannakakis} (1993).
The travelling salesman problem with distances one and two. 
\emph{Mathematics of Operations Research 18}, 1--11.

\bibitem{Zenklusen}
{\sc R. Zenklusen} (2019).
A 1.5-Approximation for path TSP. 
\emph{Proceedings of the Thirtieth Annual ACM-SIAM Symposium on Discrete Algorithms (SODA-2019)}, 1539--1549.
(see also arXiv:1805.04131v2[cs.DM])

\end{thebibliography}
\end{document}